\documentclass[sigconf]{acmart}
\AtBeginDocument{%
  }

\setcopyright{acmlicensed}
\copyrightyear{2018}
\acmYear{2018}
\acmDOI{XXXXXXX.XXXXXXX}
\acmConference[Conference acronym 'XX]{Make sure to enter the correct
  conference title from your rights confirmation email}{June 03--05,
  2018}{Woodstock, NY}
\acmISBN{978-1-4503-XXXX-X/2018/06}

\usepackage{xspace}
\usepackage{multirow}
\usepackage{tabularx}
\usepackage{geometry}

\usepackage{booktabs, tabularx, makecell}
\usepackage{enumitem}
\usepackage{xcolor}  

\newcommand{\tool}{C2RustXW\xspace}
\newcommand{\pltranslation}{PLTranslation\xspace}

\newcommand{\ctorust}{C2Rust\xspace}
\newcommand{\genctorust}{GenC2Rust\xspace}
\newcommand{\lactor}{LAC2R\xspace}
\newcommand{\verttool}{VERT\xspace}
\newcommand{\spectra}{SPECTRA\xspace}
\newcommand{\sactor}{SACTOR\xspace}
\newcommand{\saferrust}{C2SaferRust\xspace}

\begin{document}

\title{C2RustXW: Program-Structure-Aware C-to-Rust Translation via Program Analysis and LLM}

\author{Yanyan Yan}
\email{yanyanyan@smail.nju.edu.cn}
\affiliation{%
  \institution{State Key Laboratory for Novel Software Technology, Nanjing University}
  \city{Nanjing}
  \state{JiangSu}
  \country{China}
}

\author{Yang Feng}
\authornote{Yang Feng is the corresponding author.}
\affiliation{%
  \institution{State Key Laboratory for Novel Software Technology, Nanjing University}
  \city{Nanjing}
  \country{China}}
\email{fengyang@nju.edu.cn}

\author{Jiangshan Liu}
\affiliation{%
  \institution{State Key Laboratory for Novel Software Technology, Nanjing University}
  \country{China}}
\email{jiangshanl@smail.nju.edu.cn}

\author{Di Liu}
\affiliation{%
  \institution{Jiangsu Police Institute}
  \country{China}}
\email{diliujspi@gmail.com}

\author{Zixi Liu}
\affiliation{%
  \institution{State Key Laboratory for Novel Software Technology, Nanjing University}
  \country{China}}
\email{zxliu@smail.nju.edu.cn}

\author{Hao Teng}
\affiliation{%
  \institution{State Key Laboratory for Novel Software Technology, Nanjing University}
  \country{China}}
\email{221250008@smail.nju.edu.cn}

\author{Baowen Xu}
\affiliation{%
 \institution{State Key Laboratory for Novel Software Technology, Nanjing University}
 \country{China}}
\email{bwxu@nju.edu.cn}

\renewcommand{\shortauthors}{Trovato et al.}

\begin{abstract}
The growing adoption of Rust for its memory safety and performance has increased the demand for effective migration of legacy C codebases. However, existing rule-based translators (e.g., \ctorust) often generate verbose, non-idiomatic code that preserves unsafe C semantics, limiting readability, maintainability, and practical adoption. Moreover, manual post-processing of such outputs is labor-intensive and rarely yields high-quality Rust code, posing a significant barrier to large-scale migration.

To address these limitations, we present \tool, a program-structure-aware C-to-Rust translation approach that integrates program analysis with Large Language Models (LLMs). 
\tool extracts the multi-level program structure, including global symbols, function dependencies, and control- and data-flow information, and encodes these as structured textual representations injected into LLM prompts to guide translation and repair. 
Based on this design, \tool performs dependency-aware translation and adopts a multi-stage repair pipeline that combines rule-based and structure-guided LLM-based techniques to ensure syntactic correctness. 
For semantic correctness, \tool further integrates execution-based validation with structure-guided reasoning to localize and repair behavioral inconsistencies.
Experimental results show that \tool achieves 100\% syntactic correctness on CodeNet and 97.78\% on GitHub, while significantly reducing code size (up to 43.70\%) and unsafe usage (to 5.75\%). 
At the project level, \tool achieves perfect syntactic correctness and an average semantic correctness of 78.87\%, demonstrating its effectiveness for practical and scalable C-to-Rust migration.
\end{abstract}

%
%



\maketitle

\section{Introduction}
Rust has gained significant adoption due to its strong memory-safety guarantees and competitive performance, making it an appealing target for migrating legacy C codebases prone to memory errors and undefined behaviors. However, the scale of existing C systems and the high cost of manual rewriting pose substantial barriers to practical migration. Automated translation tools such as \ctorust~\cite{c2rust} provide an initial pathway by converting C code into Rust while preserving original semantics, but their outputs are often non-idiomatic and heavily rely on \texttt{unsafe} constructs. As a result, these translations typically require extensive manual refactoring to achieve safe, maintainable Rust code, thereby limiting their usability in real-world engineering workflows.

Existing C-to-Rust translation approaches face fundamental limitations. Manual translation is costly and error-prone, requiring experience in both languages, and even idiomatic Rust code still contains 20–30\% \texttt{unsafe} usage.~\cite{memory-thread-safety-practices-rust, safe-systems-programming-rust}. Automated pipelines such as \ctorust generate largely unsafe Rust code and often increase code size~\cite{empirical-rust}, with over 90\% of functions remaining unsafe~\cite{unsafe-transpiler}, thus affecting Rust’s safety goals~\cite{translating-c-rust}. Learning-based approaches~\cite{unsupervised-translation} are limited by the scarcity of high-quality parallel C–Rust corpora and high training costs. Recent LLM-based methods attempt direct translation, but still achieve limited success rates (e.g., below 40\% on real-world projects~\cite{translating-real-world-code-with-llms}). These challenges highlight the need for a translation approach that can simultaneously ensure structural fidelity, syntactic correctness, and semantic safety.

In this paper, we present \tool, a program-structure-aware C-to-Rust translation approach that integrates static program analysis with LLM-based generation and repair. 
Unlike previous approaches that rely on implicit context, \tool explicitly models the program structure, including global symbols, function dependencies, and control- and data-flow information, and transforms these structures into textual representations that are injected into LLM prompts. 
This design enables dependency-aware translation with consistent interfaces and structure-guided repair through a multi-stage pipeline combining rule-based and LLM-based techniques. 
To ensure semantic correctness, \tool further incorporates differential testing and structure-guided reasoning using control- and data-flow information and runtime states. 
For multi-file C projects, \tool extends this design by modeling inter-file dependencies and constructing a unified structural representation, enabling consistent and scalable translation.
We evaluate \tool on benchmark datasets and real-world C projects. 
At the file level, \tool achieves 100\% syntactic correctness on CodeNet and 97.78\% on GitHub, with high semantic accuracy. 
At the project level, \tool achieves perfect syntactic correctness and an average semantic correctness of 78.87\%. 
In addition, \tool significantly reduces code size (up to 49.60\%) and unsafe usage. 
These results demonstrate that \tool can produce concise, safe, and semantically consistent Rust code for both file-level and project-level translation.

Our contributions are summarized as follows:
\begin{itemize}[leftmargin=*]

\item \textbf{A program-structure-aware translation framework.}
We propose \tool, a C-to-Rust translation framework that explicitly models multi-level program structure, including global symbols, function dependencies, and control- and data-flow information. 
\tool systematically transforms these structures into textual representations and injects them into LLM prompts, enabling structure-aware translation that preserves dependency consistency and extends from single files to multi-file projects.

\item \textbf{A structure-guided translation and repair pipeline.}
We design a unified pipeline that integrates structure-aware translation with multi-stage repair. 
The pipeline combines deterministic rule-based fixing with LLM-based repair at the function and item levels, where program structure is explicitly encoded in prompts to guide consistent, structure-aware fixing. 
For semantic correctness, \tool uses differential testing with structure-guided reasoning, integrating control- and data-flow information and runtime states to repair behavioral inconsistencies.

\item \textbf{Comprehensive evaluation on benchmarks and projects.}
We evaluate \tool on CodeNet, a GitHub dataset, and multiple real-world C projects. 
The results show that \tool achieves up to 100\% syntactic correctness and strong semantic correctness, while significantly reducing code size and unsafe usage compared to existing approaches. 
These results demonstrate the effectiveness of \tool for practical and scalable C-to-Rust migration.

\end{itemize}

\section{Background}

\subsection{Rust Overview}

Rust has gained widespread adoption due to its strong memory-safety guarantees and competitive performance, making it a promising alternative to C/C++ for systems programming. 
Its safety model is built on ownership, borrowing, and lifetimes, which enforce memory safety at compile time and eliminate common errors such as null dereferences, data races, and memory leaks without relying on garbage collection. 
This design enables efficient and concurrent systems while retaining fine-grained control over hardware resources. 
However, migrating C/C++ codebases to Rust remains challenging due to fundamental differences in memory management, type systems, and abstraction mechanisms. 
Although Rust supports interoperability with C/C++ through foreign function interfaces (FFI) and provides \texttt{unsafe} constructs for low-level operations, excessive reliance on \texttt{unsafe} undermines Rust’s safety guarantees and reduces maintainability. 
Therefore, achieving semantic preservation while minimizing unsafe usage is a central challenge.

\subsection{C-to-Rust Challenges}

Translating legacy C code to Rust remains challenging due to fundamental differences in memory safety models and language abstractions. 
Existing rule-based tools such as \ctorust~\cite{transpiler-safer-rust} provide an initial solution by preserving C semantics, but rely heavily on \texttt{unsafe} constructs and FFIs, which bypass Rust’s safety guarantees. 
Prior work shows that such translations can significantly inflate code size and produce code that is almost entirely wrapped in \texttt{unsafe}~\cite{empirical-rust}, undermining Rust’s core advantages. 
Furthermore, the generated code often mixes C-style constructs, low-level types (e.g., \texttt{libc::c\_int}), and external interfaces (e.g., \texttt{malloc}), resulting in verbose and non-idiomatic implementations that are difficult to maintain. 
Handling complex C features, such as macros and low-level system interactions, further complicates translation~\cite{c2rustdoc}, and even when successful, the resulting code often requires substantial manual refinement. 
These limitations highlight the need for approaches that not only preserve semantics but also improve safety, code quality, and structural consistency, motivating the integration of program analysis with LLM-based techniques.

\begin{table*}[htbp]
  \centering
  \footnotesize
  \caption{Approaches and their features for C-to-Rust translation}
    \begin{tabular}{|c|rc|cr|cr|rcr|cr|c|}
    \toprule
    \multirow{2}[3]{*}{Baseline} & \multicolumn{2}{c|}{Methodology} & \multicolumn{2}{c|}{Syntax fixing} & \multicolumn{2}{c|}{Semantic fixing} & \multicolumn{3}{c|}{Granularity} & \multicolumn{2}{c|}{Mode} & \multirow{2}[3]{*}{OpenSource} \\
\cmidrule{2-12}          & \multicolumn{1}{c|}{Rules} & LLM   & \multicolumn{1}{c|}{LLM} & \multicolumn{1}{c|}{Rules} & \multicolumn{1}{c|}{LLM} & \multicolumn{1}{c|}{Rules} & \multicolumn{1}{c|}{function} & \multicolumn{1}{c|}{file} & \multicolumn{1}{c|}{project} & \multicolumn{1}{c|}{Transpile} & \multicolumn{1}{c|}{Refactor} &  \\
    \midrule
    C2RustXW & \multicolumn{1}{c}{\checkmark} & \checkmark     & \checkmark     & \multicolumn{1}{c|}{\checkmark} & \checkmark     & \multicolumn{1}{c|}{\checkmark} & \multicolumn{1}{c}{\checkmark} & \checkmark     & \multicolumn{1}{c|}{\checkmark} & \checkmark     &       & \checkmark \\
    C2Rust & \multicolumn{1}{c}{\checkmark} &       &       &       &       &       &       & \checkmark     & \multicolumn{1}{c|}{\checkmark} & \checkmark     &       & \checkmark \\
    C2SaferRust & \multicolumn{1}{c}{\checkmark} & \checkmark     & \checkmark     &       & \checkmark     &       &       &       & \multicolumn{1}{c|}{\checkmark} &       & \multicolumn{1}{c|}{\checkmark} & \checkmark \\
    GenC2Rust & \multicolumn{1}{c}{\checkmark} &       &       &       &       &       &       & \checkmark     & \multicolumn{1}{c|}{\checkmark} & \checkmark     &       & \checkmark \\
    PLTranslation &       & \checkmark     & \checkmark     &       &       &       &       & \checkmark     &       & \checkmark     &       & \checkmark \\
    SACTOR & \multicolumn{1}{c}{\checkmark} & \checkmark     & \checkmark     &       & \checkmark     &       &       & \checkmark     &       & \checkmark     &       & \checkmark \\
    LAC2R &       & \checkmark     & \checkmark     &       & \checkmark     &       &       & \checkmark     & \multicolumn{1}{c|}{\checkmark} &       & \multicolumn{1}{c|}{\checkmark} & Closed Source \\
    VERT  &       & \checkmark     &       &       &       &       &       & \checkmark     &       & \checkmark     &       & \checkmark \\
    SPECTRA &       & \checkmark     & \checkmark     &       & \checkmark     &       &       & \checkmark     &       & \checkmark     &       & Invalid Link \\
    \bottomrule
    \end{tabular}%
  \label{tab: c2rust-baseline}%
\end{table*}%

\subsection{C-to-Rust Translation Approaches}

Existing C-to-Rust translation approaches can be broadly categorized into rule-based, refactoring-based, LLM-based, and agent-based methods, as summarized in Table~\ref{tab: c2rust-baseline}. 
Rule-based approaches, such as \ctorust~\cite{c2rust} and \genctorust~\cite{genc2rust}, rely on predefined transformations and preserve C semantics, but generate verbose, non-idiomatic Rust code with extensive use of \texttt{unsafe} and limited support for semantic repair. 
Refactoring-based approaches~\cite{translating-c-rust, replacing-output-parameters-with-algebraic-data-types, c2saferrust, pr2, llm-multi-step} attempt to improve safety through multi-stage rewriting or LLM-assisted transformations, but depend on unsafe intermediate representations and incur additional transformation cost. 
LLM-based approaches~\cite{rustmap, study-introduce-llm-translating-code, spectra, vert} directly generate Rust code and can produce more idiomatic outputs, but often lack structural consistency, leading to unstable syntax correctness and incorrect dependencies, especially for large or multi-file programs. 
Agent-based approaches~\cite{LLM-agent, transagent} improve repair via iteration, but introduce system complexity and scalability challenges.

Despite these advances, existing approaches remain fundamentally limited by the lack of explicit modeling of program structure. 
Rule-based methods over-preserve low-level semantics, refactoring-based approaches struggle to recover from structurally inconsistent intermediate code, and LLM-based methods rely on implicit context that is insufficient for maintaining interface and dependency consistency. 
In contrast, \tool adopts a program-structure-aware design that explicitly models global symbols and dependency relationships and incorporates them into both translation and repair. 
This enables dependency-aware code generation, structure-guided syntax fixing, and semantic repair, allowing \tool to support function-, file-, and project-level translation within a unified framework while producing safer, more concise, and more consistent Rust code than existing approaches.

\section{Approach}

\begin{figure*}[htbp]
  \centering
  \includegraphics[]{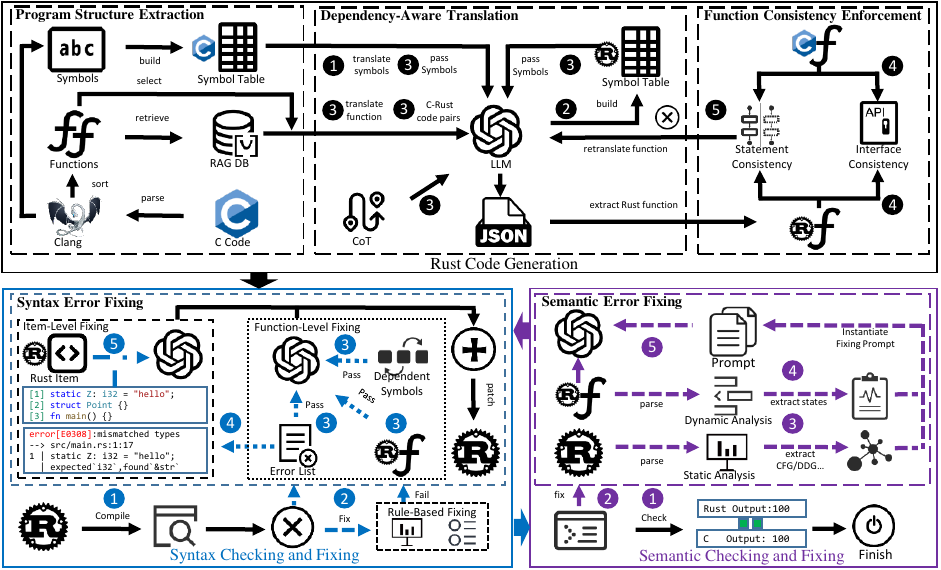}
  \caption{The overall architecture design of \tool.}
  \Description{This is the picture of the Rust code generation framework of \tool, consisting of data preparation, code translation, and function consistency checking.}
  \label{fig: framework}
\end{figure*}

\tool is a program-structure-aware framework that translates C programs into safe, compilable Rust. It models program structure as global symbols, function interfaces, and inter-function dependencies, extracted via static analysis (e.g., symbol tables and call graphs). This structure guides both translation and repair processes. \tool consists of three stages. \textit{Rust Code Generation} translates symbols and functions in topological order, ensuring dependency-aware translation. \textit{Syntax Checking and Fixing} resolves errors through rule-based and structure-aware LLM repair. \textit{Semantic Checking and Fixing} ensures behavioral equivalence with differential testing and structure-guided repair. Figure~\ref{fig: framework} shows that program structure is the central abstraction.

\begin{table}[t]
  \footnotesize
  \centering
  \caption{Program structure extraction, representation, and usage in \tool}
  \setlength{\tabcolsep}{3pt}
  \begin{tabularx}{\columnwidth}{@{}p{0.15\columnwidth} p{0.23\columnwidth} p{0.18\columnwidth} X p{0.08\columnwidth}@{}}
    \toprule
    \textbf{Structure} & \textbf{Extraction \& contents} & \textbf{Storage (example)} & \textbf{LLM representation \& usage} & \textbf{Stage} \\
    \midrule

    Global \newline symbols ($\Sigma$)
    & Clang/syn AST; \newline global vars, \newline struct/enum, function signatures
    & dict: \newline\texttt{\{g:int,\; sum:fn\}}
    & Retrieved via dependency names and injected as \texttt{dependent symbols} for \newline type/interface context
    & T, S \\

    Functions ($\mathcal{F}$)
    & AST nodes; signature + body (or summary)
    & obj: \newline\texttt{fn sum(a,b)}
    & Used as main translation/repair unit with interface constraints preserved
    & T, S \\

    Dependencies ($\mathcal{D}$)
    & AST traversal; symbol usage and calls
    & list: \texttt{[g, Node]}
    & Expanded via $\Sigma$ \newline lookup into \newline dependency-aware prompt context
    & T, S \\

    Call graph
    & Function call analysis; DAG
    & graph: \texttt{main$\rightarrow$foo}
    & Determines translation order (callee before caller)
    & T \\

    Rust item \newline structure
    & syn AST; \newline struct/impl/trait
    & AST node: \texttt{struct Node}
    & Extracted as repair scope for item-level fixing
    & S \\

    CFG
    & Static analysis; basic blocks + edges
    & blocks: \texttt{B1$\rightarrow$B2}
    & Serialized into textual control-flow blocks for reasoning
    & M \\

    DDG
    & Def-use and variable dependencies
    & edges: \texttt{x$\rightarrow$y}
    & Converted to dependency chains to expose data flow
    & M \\

    Runtime state
    & Instrumentation; execution traces
    & map: \texttt{\{x=5,y=10\}}
    & Injected as variable states to localize semantic inconsistencies
    & M \\

    File-level \newline dependencies
    & cross-file analysis and symbol usage
    & graph and symbol table
    & Guides file translation order and provides unified symbol context across files
    & T \\

    \bottomrule
  \end{tabularx}
  \parbox{\columnwidth}{\footnotesize
  \textit{Note:} T = Translation; S = Syntax fixing; M = Semantic fixing.}
  \label{tab: structure-all}
\end{table}

\subsection{Structure Definition and Representation}

A key concept in \tool is \emph{program structure}, which is explicitly modeled and propagated throughout the translation and repair pipeline. 
Given a C program $P$, we represent its program structure as a tuple $\mathcal{S}(P) = (\Sigma, \mathcal{F}, \mathcal{D}, \mathcal{G})$, where $\Sigma$ denotes the global symbol table (including types, variables, structs, and function signatures), $\mathcal{F}$ represents functions with their interfaces and bodies, $\mathcal{D}$ captures dependency relations among program elements (e.g., function calls and symbol usage), and $\mathcal{G}$ encodes control- and data-flow structures (e.g., CFG and DDG). 
This representation is extracted via static analysis and serves as a unified abstraction of program semantics.

To construct $\mathcal{S}(P)$, \tool performs structure extraction and representation for both C and Rust programs. 
For C code, we use \texttt{Clang/LLVM} to parse the program into an AST, from which global symbols, function definitions, and dependency relations are extracted by traversing AST nodes. 
Global symbols are stored in a dictionary mapping identifiers to their definitions, while function-level dependencies are recorded as symbol lists associated with each function. 
Control-flow and data-flow structures are obtained via static analysis and represented as node--edge relations (e.g., CFG and DDG). 
For Rust code, we use the \texttt{syn} parser to construct ASTs, from which function bodies, items (e.g., structs, traits, and impls), and symbol information are extracted for subsequent repair and analysis. 
Across both languages, these structures are stored in lightweight representations such as dictionaries, lists, and graphs that can be efficiently queried and incrementally updated.

To enable LLM-based translation and repair, all program structures are further serialized into textual form and structured into prompts. 
For example, given a function to be translated or repaired, \tool first retrieves its dependency list from $\mathcal{D}$, then queries $\Sigma$ to obtain the corresponding symbol definitions, and finally inserts these definitions into the \texttt{dependent symbols} section of the prompt, with one definition per line. 
Similarly, control-flow and data-flow structures are converted into textual node-edge descriptions, and runtime states are represented as key-value mappings. 
This design bridges structured program analysis and LLM reasoning, allowing the model to operate on explicit program structure rather than implicit context.

The program structure is not only used to describe the input program but also serves as a constraint and guidance throughout all stages of \tool. 
During translation, it determines the order of generation and provides dependency-aware context for each function; during syntax fixing, it constrains repair within well-defined structural scopes such as statements, functions, and items; and during semantic fixing, it enables structure-guided reasoning by incorporating control-flow and data-flow information. 
By explicitly modeling and preserving $\mathcal{S}(P)$, \tool ensures that both generated and repaired Rust code remain consistent with the original program’s structural and semantic intent.
This structure-aware approach minimizes semantic drift, ensuring the final Rust program accurately reflects the original C program's behavior.

\subsection{Rust Code Generation}

The Rust code generation stage translates C programs into Rust in a program-structure-aware manner, incorporating extracted structural information into the translation process. 
Instead of monolithic generation, \tool performs dependency-aware translation by organizing code into structured units and injecting program structure into LLM prompts, ensuring generated code respects symbol dependencies, function interfaces, and program semantics.

\noindent\textbf{Dependency-Aware Translation.}
\tool performs translation in two phases. 
First, global symbols are translated into Rust to establish a consistent interface foundation. 
Then, functions are translated in dependency order, determined by the call graph, ensuring that callees are defined before callers. 
For each function, \tool retrieves its dependency list from $\mathcal{D}$ and queries the global symbol table $\Sigma$ to obtain the corresponding symbol definitions. 
These definitions, including global variables, type declarations, and function signatures, are serialized into textual form and inserted into the \texttt{dependent symbols} section of the prompt. 
As shown in Table~\ref{tab: unified-prompt-template}, the prompt for each function consists of the original C function, its dependent symbol definitions (in both C and Rust forms), and additional constraints, enabling the LLM to generate translations with full structural context. 
This prompt construction ensures the generated Rust code maintains consistent interfaces and avoids mismatched symbol definitions.

\noindent\textbf{Project-Level Dependency Handling.}
For multi-file C projects, \tool extends this strategy by incorporating inter-file dependencies into the translation process. 
A file-level dependency graph is constructed based on include relations and cross-file symbol usage, and a topological order is used to determine the translation sequence. 
All source files are then unified into a consistent structural representation with deduplicated symbols, forming a project-level extension of $\mathcal{S}(P)$ that enables translation to proceed in a structure-aware manner across files.
allowing the translation process to proceed in the same manner as single-file translation while preserving cross-file consistency. 
By explicitly injecting dependency information into prompts, \tool ensures that symbols defined in one file are correctly referenced in other files, avoiding inconsistencies that commonly arise during project translation.

\noindent\textbf{Function Consistency Enforcement.}
After initial translation, \tool enforces consistency between C and Rust functions to ensure interface and structural alignment. 
This includes validating function signatures against the global symbol table, removing invalid or hallucinated constructs, and checking structural correspondence between C and Rust function bodies. 
Rather than requiring exact syntactic equivalence, \tool compares normalized statement categories (e.g., control flow, assignments, and declarations) to preserve the intent of control-flow and data-flow. 
If inconsistencies are detected, functions are selectively re-translated under refined structural constraints. 
Finally, translated global symbols and functions are integrated into a complete Rust program by removing duplicate definitions and ordering items according to dependency constraints, producing a coherent and compilable output. 
For project-level translation, outputs from all source files are unified into a single representation to ensure consistency across subsequent stages of syntax and semantic fixing.

\begin{table*}[htbp]
  \footnotesize
  \centering
  \caption{Unified prompt templates for function translation, compilation error fixing, and semantic error fixing}
  \renewcommand{\arraystretch}{0.9}
  \resizebox{\linewidth}{!}{
  \begin{tabular}{|l|l|l|l|l|}
    \toprule
    \multicolumn{1}{|c|}{\textbf{Dependancy-aware translation}} 
    & \multicolumn{2}{c|}{\textbf{Syntax error fixing}} 
    & \multicolumn{2}{c|}{\textbf{Semantic error fixing}} \\
    \midrule

    \textbf{Translation instruction} 
    & \textbf{Function-level fixing} 
    & \textbf{Item-level fixing} 
    & \textbf{Textual CFG/DDG} 
    & \textbf{Semantic fixing} \\
    
    \midrule

    \textbf{ith C function:} 
    & \textbf{Guidelines} 
    & \textbf{Guidelines} 
    & \textbf{CFG nodes} 
    & \textbf{C-Rust structure information} \\

    <C function code> 
    & \textbf<Error list>
    & <Single Error>
    & \textbf{Block i: <block code>} 
    & <C-Rust input/output> \\

    \textbf{Dependent symbols} 
    & \textbf{Selected Rust function}
    & \textbf{Selected item} 
    &  \textbf{CFG edges}
    & <C/Rust output diff> \\

    <C symbol definitions> 
    & <Rust function code>
    & <Location> 
    & <Block i -> Block j>
    & <C/Rust output code> \\

    <Rust symbol definitions> 
    & <Function interface>
    & <Code> 
    & \textbf{DDG nodes}
    & <C CFG/DDG information> \\

    \textbf{RAG code pairs} 
    & \textbf{Dependent symbols}
    & \textbf{Dependent symbols} 
    & <Node i (symbol)>
    & \textbf{Instrumented runtime states}\\

    <C-Rust code pairs> 
    &  <Rust symbol definitions>
    & <Rust symbol definitions> 
    & \textbf{DDG edges} 
    & <identifier -> value> \\

    \textbf{Constraints} 
    & \textbf{Constraints}
    & \textbf{Constraints} 
    & <Node i -> Node j> 
    & <Code with semantic errors> \\

    \bottomrule
  \end{tabular}
  }
  \label{tab: unified-prompt-template}
\end{table*}

\subsection{Syntax checking and fixing}

After code generation, \tool ensures that the translated Rust program is syntactically valid through a multi-stage repair pipeline. 
Syntax checking is performed by compiling the generated code using \texttt{cargo}, which produces structured diagnostic messages with precise locations and error categories. 
These diagnostics not only identify syntax violations but also expose deeper inconsistencies such as type mismatches, unresolved symbols, and ownership-related issues introduced during translation. 
Given the heterogeneous nature of these failures, \tool organizes syntax fixing into a hybrid framework applied in order of increasing structural scope, where program structure (Table~\ref{tab: structure-all}) is progressively incorporated into repair contexts and serialized into prompts (Table~\ref{tab: unified-prompt-template}).

\noindent\textbf{Unparsable code fixing.}
Some generated Rust code cannot be parsed due to severe syntax errors such as missing delimiters or malformed expressions, preventing subsequent analysis. 
To address this, \tool first performs coarse-grained repair by extracting a local code region around the error location and prompting the LLM to rewrite this region. 
This step restores syntactic parseability without relying on structured context, enabling subsequent stages to operate on well-formed code.

\noindent\textbf{Rule-based Statement-level Fixing.}
Once the code becomes parsable, \tool applies deterministic rule-based transformations derived from static analysis. 
The Rust code is parsed into an AST, and compiler diagnostics are matched to predefined error patterns, enabling targeted corrections at the statement level. 
Representative rules include: (1) symbol resolution using the global symbol table $\Sigma$, (2) type correction to satisfy Rust’s strict typing constraints, (3) import completion and deduplication, (4) variable normalization to resolve naming conflicts, and (5) insertion of unsafe contexts when required. 
These transformations directly modify the AST and regenerate code, ensuring high precision and minimal edits. 
By resolving frequent and well-structured errors without LLM intervention, this stage significantly reduces the complexity of subsequent structure-aware repair.

\noindent\textbf{Dependency-Aware Function-Level Fixing.}
For errors involving function bodies and their interactions with global symbols, \tool performs function-level repair guided by program structure. 
For each function, we construct a structured repair context consisting of its definition, interface, error messages, and dependent symbols retrieved from $\Sigma$ (Table~\ref{tab: structure-all}). 
These elements are serialized into the function-level fixing prompt (Table~\ref{tab: unified-prompt-template}, second column), where dependent symbol definitions are explicitly listed to provide type and interface context. 
This dependency-aware prompt enables the LLM to repair multiple errors within the function while preserving structural consistency.

\noindent\textbf{Structure-Aware Item-Level Fixing.}
Remaining errors that span beyond individual functions or involve higher-level constructs (e.g., type definitions or module-level inconsistencies) are handled at the item level. 
For each error, \tool identifies the minimal enclosing Rust item and extracts its definition along with relevant structural context (e.g., referenced symbols from $\Sigma$). 
This information is serialized into the item-level fixing prompt (Table~\ref{tab: unified-prompt-template}, third column), allowing the LLM to perform repair under explicit structural constraints. 
To improve reliability, \tool requires the LLM to return both original and modified code, ensuring consistency and preventing unintended changes.
These fixing techniques are applied iteratively until compilation succeeds or a limit is reached.

\subsection{Semantic checking and fixing}
While syntactic correctness ensures compilability, it does not guarantee semantic correctness. 
\tool therefore performs semantic checking and fixing to ensure that the translated Rust program faithfully preserves the behavior of the C program. 
As illustrated in Fig.~\ref{fig: framework}, this stage follows a closed-loop workflow that integrates differential testing, program structure (Table~\ref{tab: structure-all}), and LLM-guided repair via structured prompts (Table~\ref{tab: unified-prompt-template}).

\noindent\textbf{Semantic consistency checking.}
Semantic checking relies on executable C programs that contain a \texttt{main} function serving as an execution driver in \tool. 
This design ensures that each input program can be compiled and executed independently, enabling direct behavioral comparison. 
Given such a driver, we execute both the C program and its translated Rust counterpart under identical test inputs and compare their outputs. 
All programs are associated with predefined test cases, which are reused to drive execution and ensure consistent evaluation conditions. 
By treating the C program as the reference implementation, differential testing detects semantic inconsistencies whenever the Rust output deviates from the expected behavior. 
This execution-based validation provides a reliable and scalable mechanism for identifying semantic errors that cannot be captured through static checks alone.

\noindent\textbf{Semantic Error Localization and Analysis.}
When inconsistencies are detected, \tool performs structure-aware analysis to localize and characterize semantic errors. 
First, output differences between C and Rust executions are computed to identify behavioral mismatches, and output-related statements are used to associate discrepancies with corresponding code regions. 
To enable deeper reasoning, \tool retrieves control-flow and data-flow structures (CFG and DDG) from the program structure (Table~\ref{tab: structure-all}) and serializes them into textual representations. 
These representations are injected into the fixing prompt (Table~\ref{tab: unified-prompt-template}, fourth column), allowing the LLM to reason about execution paths and data dependencies beyond surface syntax.
In addition to static structure, \tool incorporates dynamic program structure by injecting runtime states collected from instrumented Rust execution (Table~\ref{tab: structure-all}). 
These states are represented as key-value mappings of intermediate variables and provide fine-grained behavioral evidence, enabling the tracing of discrepancies across both control and data-flow dimensions.

\noindent\textbf{Structure-Guided Semantic Fixing.}
Based on the combined structural context, \tool constructs a semantic repair prompt to guide the LLM in correcting behavioral inconsistencies. 
The prompt (Table~\ref{tab: unified-prompt-template}, fifth column) integrates: (1) C and Rust inputs and outputs, (2) output differences, (3) output-related code statements, (4) textual CFG/DDG representations, and (5) runtime states. 
This unified representation explicitly encodes both intended behavior (from C) and observed deviations (from Rust), grounded in program structure. The LLM is then instructed to generate revised Rust code that preserves the original semantics while satisfying Rust’s language constraints. 
The repaired code is recompiled and re-evaluated through differential testing, forming an iterative refinement loop until semantic equivalence is achieved or a predefined limit is reached. 
By combining execution-based validation with structure-guided reasoning, \tool effectively resolves semantic inconsistencies that cannot be addressed through syntax-level repair alone while maintaining structural consistency across stages.

\section{Experiment Setup}


\begin{table*}[htbp]
  \centering
  \footnotesize
  \caption{Statistics of file-level (CodeNet and GitHub) and project-level datasets.}
    \resizebox{\linewidth}{!}{
    \begin{tabular}{ccccccc|cccccccc}
    \toprule
    \multicolumn{7}{c|}{File-level datasets}              & \multicolumn{8}{c}{Project-level datasets} \\
    \midrule
    Dataset & \multicolumn{1}{c}{Category} & Files & LOCs  & Funcs & Structs & LineCov(\%) & \multicolumn{1}{c}{Project} & C Files & headers & CLOC  & Functions & Structs & Macros & Global Variables \\
\cmidrule{2-7}\cmidrule{9-15}    \multicolumn{1}{c}{\multirow{3}[2]{*}{CodeNet}} & <50   & 149   & 3260  & 183   & 2     & 94    & ht    & 2     & 1     & 246   & 10    & 3     & 4     & 0 \\
          & 50-200 & 49    & 4431  & 306   & 41    & 88    & quadtree & 6     & 1     & 496   & 31    & 4     & 3     & 0 \\
          & >200  & 2     & 827   & 58    & 20    & 55    & rgba  & 3     & 1     & 569   & 19    & 2     & 3     & 0 \\
\cmidrule{2-7}    \multicolumn{1}{c}{\multirow{3}[2]{*}{GitHub}} & <50   & 28    & 1087  & 56    & 0     & 96.38 & urlparser & 2     & 1     & 634   & 21    & 1     & 6     & 1 \\
          & 50-200 & 50    & 4371  & 186   & 6     & 95.18 & genann & 4     & 2     & 877   & 22    & 1     & 10    & 2 \\
          & >200  & 12    & 3303  & 106   & 15    & 83.51 & libcsv & 3     & 1     & 1115  & 31    & 2     & 35    & 1 \\
    \bottomrule
    \end{tabular}%
    }
  \label{tab: dataset}
\end{table*}

\begin{table*}[htbp]
  \centering
  \footnotesize
  \caption{File-level effectiveness and code quality comparison on CodeNet and GitHub datasets}
  \renewcommand{\arraystretch}{0.8}
  \begin{tabularx}{\textwidth}{
c 
p{0.06\textwidth} |
p{0.09\textwidth} p{0.09\textwidth} p{0.10\textwidth} p{0.05\textwidth} |
p{0.09\textwidth} p{0.09\textwidth} p{0.10\textwidth} p{0.05\textwidth}
}
    \toprule
    \multirow{2}{*}{Approach} & \multirow{2}{*}{Category}
    & \multicolumn{4}{c|}{CodeNet}
    & \multicolumn{4}{c}{GitHub} \\
    \cmidrule{3-10}
    & & SynCor$\uparrow$ & SemCor$\uparrow$ & RLOC$\downarrow$/PUR$\downarrow$ & \#W$\downarrow$/\#E$\downarrow$
      & SynCor$\uparrow$ & SemCor$\uparrow$ & RLOC$\downarrow$/PUR$\downarrow$ & \#W$\downarrow$/\#E$\downarrow$ \\
    \midrule

    \multirow{4}{*}{C2Rust}
    & <50   & 147(99.32) & 144(97.96) & 8485/70.12 & 761/5 & \textbf{28(100.00)} & \textbf{27}(96.43) & 1797/72.62 & 186/\textbf{0} \\
    & 50-200& \textbf{49(100.00)} & 47(95.92)     & \textcolor{red}{375147}/30.96 & 972/8 & 49(98.00) & \textbf{49(100.00)} & \textcolor{red}{1007082/99.88} & 787/\textbf{0} \\
    & >200  & \textbf{2(100.00)} & 2(100.00)         & 1543/84.71 & 161/0 & \textbf{11(91.67)} & \textbf{9}(81.82) & 3622/79.60 & 342/1 \\
    & Total & 198(99.00) & 193(97.47) & \textcolor{red}{385175}/32.04 & 1894/13 & \textbf{88(97.78)} & \textbf{85}(96.59) & \textcolor{red}{1012501/99.76} & 1315/\textbf{1} \\

    \multirow{4}{*}{PLTranslation}
    & <50   & 146(98.65) & 62(42.47)  & 4178/1.82 & 89/4 & \textbf{28(100.00)} & 24(85.71) & 1031/\textbf{0.00} & 17/\textbf{0} \\
    & 50-200& 36(73.47) & 9(25.00)    & \textbf{2862}/\textbf{8.77} & \textbf{107}/7 & 38(77.55) & 35(92.11) & 2576/\textbf{0.00} & 32/2 \\
    & >200  & 0(0.00) & 0(0.00)                   & 0/0.00 & 0/0 & 8(72.73)  & 3(37.50) & 1385/10.25 & 17/\textbf{0} \\
    & Total & 182(91.00) & 71(39.01)  & 7040/4.64 & 196/11 & 74(82.22) & 62(83.78) & 4992/\textbf{2.84} & 66/2 \\

    \multirow{4}{*}{GenC2Rust}
    & <50   & 147(99.32) & 144(97.96) & 7317/81.43 & 772/4 & \textbf{28(100.00)} & \textbf{27}(96.43) & 1576/82.99 & 187/\textbf{0} \\
    & 50-200& \textbf{49(100.00)} & 48(97.96)     & \textcolor{red}{374759}/30.99 & 996/8 & 49(98.00) & \textbf{49(100.00)} & \textcolor{red}{1006578/99.92} & 789/\textbf{0} \\
    & >200  & \textbf{2(100.00)} & \textbf{2(100.00)}         & 1513/85.46 & 188/0 & \textbf{11(91.67)} & \textbf{9}(81.82) & 3492/81.36 & 354/1 \\
    & Total & 198(99.00) & \textbf{194(97.98)} & \textcolor{red}{383589/32.17} & 1956/12 & \textbf{88(97.78)} & \textbf{85}(96.59) & \textcolor{red}{1011646/99.83} & 1330/\textbf{1} \\

    \multirow{4}{*}{SACTOR}
    & <50   & 25(16.78) & 2(8.00)     & \textbf{451}/\textbf{0.00} & \textbf{2}/\textbf{0} & 4(14.29) & 4(\textbf{100.00}) & \textbf{103}/\textbf{0.00} & \textbf{1}/\textbf{0} \\
    & 50-200& 0(0.00) & 0(0.00)                   & 0/0.00 & 0/0 & 17(34.69) & 17(100.00) & \textbf{620}/16.29 & \textbf{21}/13 \\
    & >200  & 0(0.00) & 0(0.00)                   & 0/0.00 & 0/0 & 1(9.09) & 1(\textbf{100.00}) & \textbf{109}/\textbf{0.00} & \textbf{3}/\textbf{0} \\
    & Total & 25(12.50) & 2(8.00)     & \textbf{451}/\textbf{0.00} & \textbf{2}/\textbf{0} & 22(24.44) & 22(\textbf{100.00}) & \textbf{832}/12.14 & \textbf{25}/13 \\

    \midrule

    \multirow{4}{*}{C2RustXW}
    & <50   & \textbf{149(100.00)} & \textbf{145}(97.32) & 3802/9.55 & 370/4 & \textbf{28(100.00)} & \textbf{27}(96.43) & 878/0.80 & 97/\textbf{0} \\
    & 50-200& \textbf{49(100.00)} & 34(69.39)   & 4962/37.00 & 721/\textbf{0} & \textbf{50(100.00)} & \textbf{49}(98.00) & 2552/4.19 & 254/2 \\
    & >200  & \textbf{2(100.00)} & 1(50.00)     & \textbf{51/0.00} & \textbf{0/0} & 10(90.91) & 8(80.00) & 1233/12.49 & 99/\textbf{0} \\
    & Total & \textbf{200(100.00)} & 180(90.00) & 8815/24.95 & 1091/4 & \textbf{88(97.78)} & 84(95.45) & 4663/5.75 & 450/2 \\

    \bottomrule
  \end{tabularx}
  \label{tab: combined-file-results}
\end{table*}

This section outlines the research questions designed to evaluate the effectiveness of our approach and the experimental setup, including data preparation, baselines, and evaluation metrics.

\subsection{Research Questions}

We formulate three research questions to evaluate \tool across different translation granularities.

\noindent\textbf{RQ1. File-level effectiveness.}
How effectively does \tool translate individual C files into correct and high-quality Rust code?
This evaluates syntactic correctness, semantic consistency, and code quality at the file level.

\noindent\textbf{RQ2. Project-level effectiveness.}
How effectively does \tool translate complete C projects while preserving correctness and scalability?
This evaluates performance on real-world projects with inter-file dependencies.

\noindent\textbf{RQ3. Qualitative analysis.}
What insights can be gained about the strengths and limitations of \tool compared to existing approaches?
This analyzes representative cases to understand translation behavior and differences.

\subsection{Data Preparation}

We evaluate \tool on both file-level and project-level datasets to assess its effectiveness under different translation granularities. 
For file-level evaluation, we collect 200 C programs from CodeNet~\cite{codenet-dataset, codenet-paper} and 90 programs from GitHub (The-Algorithms~\cite{the-algorithms}), all of which can be compiled and executed independently. 
The programs are categorized by size (<50, 50–200, >200 LOC) as shown in Table~\ref{tab: dataset}. 
CodeNet provides standardized benchmarks with well-defined inputs and outputs, while GitHub samples introduce more realistic coding styles and structural diversity. 
For project-level evaluation, we select six representative multi-file C projects~\cite{ht-project, quadtree-project, rgba-project, urlparser-project, genann-project, libscv-project} (e.g., \texttt{genann}, \texttt{libcsv}) following prior work~\cite{Ownership-guided-C-to-Rust-translation, genc2rust}, which contain non-trivial inter-file dependencies and diverse structures.

To enable semantic evaluation, all programs are associated with predefined test cases. 
Following prior work~\cite{study-introduce-llm-translating-code}, we construct standardized input–output pairs for both file-level and project-level programs by reusing and normalizing existing test cases into a unified format. 
This allows both C and translated Rust programs to be executed under identical inputs, ensuring fair and reproducible differential testing across all approaches. 
Finally, we apply a unified preprocessing pipeline to normalize input programs, including parsing and basic code normalization, so that all methods are evaluated under consistent and comparable experimental conditions.

\subsection{Baselines}

We compare \tool with representative C-to-Rust translation approaches at both file and project levels. At the file level, we include \ctorust, \genctorust, \pltranslation, and \sactor. \ctorust is a rule-based transpiler that translates C99 code into unsafe Rust, serving as a strong baseline for syntax-preserving translation. \genctorust extends \ctorust by improving the handling of \texttt{void*} pointers. \pltranslation adopts LLM-based direct translation, representing purely generative approaches without structural constraints. \sactor combines rule-based translation and LLM-guided refactoring in a two-stage pipeline, aiming to improve safety and semantic correctness. At the project level, we evaluate \ctorust and \genctorust for end-to-end project translation, and \saferrust, which further refactors unsafe Rust into safer code using LLM-based techniques and test-driven validation. All LLM- and agent-based approaches, including baselines and \tool, use DeepSeek-V3 for fairness and consistency.

We exclude several approaches due to practical limitations. \lactor supports both file- and project-level translation, but cannot be evaluated due to the lack of publicly available implementation and documentation. \verttool relies on the Transcoder-IR dataset~\cite{transcoder-IR}, which differs significantly from real-world C code and is therefore not suitable for our setting. \spectra is also excluded due to unavailable source code. Overall, the selected baselines cover rule-based, refactoring-based, and LLM-based approaches, enabling a comprehensive and fair comparison with \tool.

\subsection{Metrics}

We evaluate translation correctness using syntactic correctness ($SynCor$) and semantic correctness ($SemCor$) at both file and project levels. 
$SynCor$ measures whether the translated Rust code can be successfully compiled using \texttt{cargo check}, while $SemCor$ measures whether the Rust code produces outputs consistent with the original C code. 
Following prior work~\cite{unsupervised-translation}, we focus on execution-based validation rather than text similarity metrics. 
Formally, $SynCor = \frac{SYN_{rs}}{N_c}$ and $SemCor = \frac{SEM_{rs}}{SYN_{rs}}$, where $N_c$ is the number of C programs, $SYN_{rs}$ is the number of syntactic Rust programs, and $SEM_{rs}$ is the number of semantically correct Rust programs.

To evaluate code quality, we consider code size, safety, and adherence to Rust best practices. 
We report the total lines of Rust code ($RLOC$) to reflect code size. 
Safety is measured using the proportion of unsafe code ($PUR = \frac{ULOC}{RLOC}$), where $ULOC$ denotes unsafe Rust lines. 
In addition, we report the number of \texttt{Clippy} warnings (\#W) and errors (\#E) as indicators of code quality. 
Only compilable Rust programs are considered in these metrics.

\section{Results}

\subsection{Answer to RQ1: File-level effectiveness}

The goal of RQ1 is to evaluate the effectiveness of \tool for file-level C-to-Rust translation, considering both correctness and code quality. We report code quality metrics only for syntactically correct programs, since semantically correct programs form a subset and exhibit similar trends. 
Table~\ref{tab: combined-file-results} provides a unified comparison across all approaches, including syntactic correctness ($SynCor$), semantic correctness ($SemCor$), code size and safety (RLOC/PUR), and \texttt{Clippy} warnings and errors on the CodeNet and GitHub datasets. 

Across both datasets, rule-based approaches such as \ctorust and \genctorust achieve consistently high $SynCor$ (around 91-100\%) and strong $SemCor$ (average exceeds 96\%) by directly preserving C semantics through construct-level translation. 
However, this correctness comes at a substantial cost. 
These methods retain low-level C abstractions via raw pointers, unsafe operations, and FFI bindings, leading to a severe code size explosion, particularly in the \texttt{50-200} category, and extensive unsafe usage. 
While FFI enables behavioral equivalence when external libraries are involved, it introduces complex hybrid builds between Rust and C libraries, increasing developer effort and deployment complexity. 
Moreover, the generated code is often non-idiomatic, as observed in RQ3, where C-style memory manipulation is preserved rather than leveraging Rust ownership abstractions. 
Hybrid approaches such as \sactor and \c2saferrust attempt to improve these outputs through LLM-based refactoring, but their effectiveness remains limited. 
Once the initial translation introduces large amounts of unsafe code and structural complexity, post-hoc rewriting becomes difficult because it must simultaneously preserve semantics, ensure compilability, and simplify the code structure. 
This leads to unstable performance, low syntactic correctness, and high refactoring cost.

In contrast, \pltranslation adopts a purely LLM-based strategy and can generate more compact, idiomatic Rust code with minimal unsafe code by leveraging learned knowledge of Rust abstractions and libraries. 
However, its performance is highly sensitive to program scale. 
While results are acceptable for small programs, translation quality degrades significantly as code size and dependency complexity increase. 
Without explicit structural guidance, the LLM must infer interfaces and dependencies from raw code text, often producing incomplete outputs, inconsistent function interactions, or unparsable code. 
Repair attempts are significantly limited by the lack of structural information, making it particularly hard to restore consistency once the code becomes structurally invalid. Both rule-based and LLM-based approaches suffer from the absence of structure-aware reasoning, either over-preserving low-level semantics or lacking sufficient consistency constraints.

In contrast, \tool achieves a more balanced trade-off between correctness, safety, and code quality. 
It consistently attains high $SynCor$ (100\% on CodeNet and 97.78\% on GitHub) and competitive $SemCor$ (90.00\% and 95.45\%, respectively), while significantly reducing both code size and unsafe usage compared to rule-based baselines. 
For example, unsafe proportions are reduced from over 70-99\% to 24.95\% on CodeNet and 5.75\% on GitHub, without the severe code expansion observed in rule-based approaches. 
These improvements stem from the program-structure-aware design of \tool, which explicitly models global symbols, function dependencies, and translation order to ensure interface consistency and avoid redundant or incorrect code generation. 
This design prevents over-preserving unsafe C semantics while maintaining structural correctness. Overall, \tool produces substantially more concise and safer Rust code by eliminating redundant symbol definitions and replacing low-level pointer manipulations with Rust-native abstractions guided by program structure. 
Compared to LLM-based approaches such as \pltranslation, which generate compact but structurally inconsistent code, \tool maintains dependency correctness and interface consistency, preventing hidden defects even when code size is slightly larger in some cases. 
Although \tool generates a moderate number of \texttt{Clippy} warnings, these are primarily stylistic (e.g., naming conventions) rather than correctness-related, and the consistently low number of \texttt{Clippy} errors further confirms the reliability of the Rust code.


\begin{table*}[htbp]
  \centering
  \footnotesize
  \caption{Project-level effectiveness and code quality comparison of six Rust projects}
  \renewcommand{\arraystretch}{0.8}
  \resizebox{\textwidth}{!}{
    \begin{tabular}{ccccccccccccc}
    \toprule
    \multirow{2}[4]{*}{Project} & \multicolumn{3}{c}{C2Rust} & \multicolumn{3}{c}{GenC2Rust} & \multicolumn{3}{c}{C2Saferrust} & \multicolumn{3}{c}{C2RustXW} \\
\cmidrule{2-13}          & SynCor$\uparrow$/SemCor$\uparrow$ & RLOC$\downarrow$/PUR$\downarrow$ & \#W$\downarrow$/\#E$\downarrow$ & SynCor$\uparrow$/SemCor$\uparrow$ & RLOC$\downarrow$/PUR$\downarrow$ & \#W$\downarrow$/\#E$\downarrow$ & SynCor$\uparrow$/SemCor$\uparrow$ & RLOC$\downarrow$/PUR$\downarrow$ & \#W$\downarrow$/\#E$\downarrow$ & SynCor$\uparrow$/SemCor$\uparrow$ & RLOC$\downarrow$/PUR$\downarrow$ & \#W$\downarrow$/\#E$\downarrow$ \\
    \midrule
    ht    & 100.00/100.00 & 265/81.51 & 41/0  & 100.00/100.00 & 251/83.67 & 41/0  & 100.00/100.00 & 304/58.22 & 10/0  & 100.00/100.00 & 200/7.50 & 20/0 \\
    quadtree & 100.00/100.00 & 1207/76.27 & 8/0   & 100.00/0.00 & 1167/79.35 & 15/0  & 100.00/19.05 & 1076/57.81 & 19/5  & 100.00/100.00 & 482/33.61 & 21/0 \\
    rgba  & 100.00/100.00 & 1852/38.66 & 9/0   & 100.00/100.00 & 1836/39.00 & 9/0   & 100.00/0.00 & 1617/21.77 & 117/1 & 100.00/100.00 & 924/0.00 & 35/1 \\
    urlparser & 100.00/100.00 & 1449/82.75 & 70/0  & 100.00/100.00 & 1372/84.18 & 109/0 & 100.00/61.36 & 1220/71.64 & 47/12 & 100.00/68.18 & 614/0.00 & 14/0 \\
    genann & 100.00/100.00 & 1863/88.19 & 114/0 & 100.00/100.00 & 1784/89.18 & 124/0 & 50.00/0.00 & 1298/52.47 & 24/0  & 100.00/88.35 & 442/20.36 & 46/0 \\
    libcsv & 100.00/100.00 & 3154/92.33 & 177/0 & 0.00/0.00 & 3201(187.09) & 1/2   & 100.00/0.00 & 2608/74.42 & 21/1  & 100.00/16.67 & 1009/6.64 & 106/0 \\
    \midrule
    Total & \textbf{100.00}/\textbf{100.00} & 9790/77.75 & 69.83/\textbf{0.00} & 83.33/66.67 & 9611/78.80 & 49.83/0.33 & 90.00/30.07 & 8123/57.21 & \textbf{39.67}/3.17 & \textbf{100.00}/78.87 & \textbf{3671}/\textbf{8.48} & 40.33/0.17 \\
    \bottomrule
    \end{tabular}%
    }
  \label{tab: combined-project-results}
\end{table*}

\begin{figure*}[ht]
  \centering
  \includegraphics[width=\linewidth]{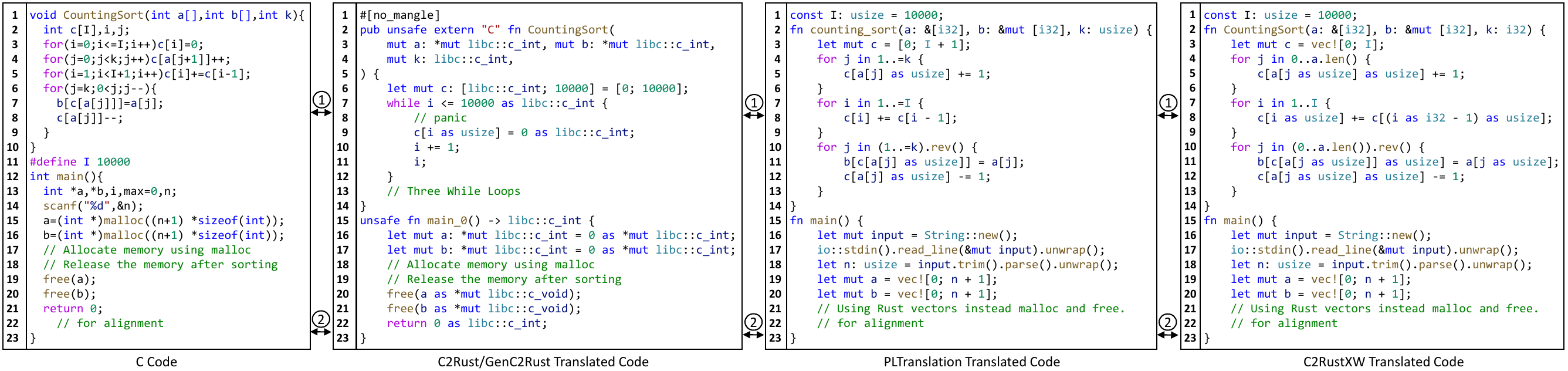}
  \caption{Counting sort program translated by \ctorust, \genctorust, \pltranslation, and \tool. (CodeNet/s787964396.c).}
  \Description{This is the counting sort program example translated by \ctorust, \genctorust, \pltranslation, and \tool.}
  \label{fig: countingsort-results}
\end{figure*}
\subsection{Answer to RQ2: Project-level effectiveness}

Table~\ref{tab: combined-project-results} presents a unified comparison of translation effectiveness and code quality across all approaches, including syntactic correctness ($SynCor$), semantic correctness ($SemCor$), code size and safety (RLOC/PUR), and \texttt{Clippy} warnings and errors. 
At the project level, translation becomes significantly more challenging than file-level scenarios, as it requires consistent handling of cross-file dependencies, module organization, and external interfaces. 
Overall, \tool achieves perfect $SynCor$ across all projects and strong $SemCor$ (78.87\% on average), demonstrating its robustness in constructing compilable and largely semantically correct Rust projects.

Rule-based approaches such as \ctorust and \genctorust also achieve perfect $SynCor$ and, in many cases, perfect $SemCor$ at the project level by preserving C semantics through FFI bindings and low-level C-style abstractions. 
However, this apparent correctness relies on a translation strategy that tightly couples the generated Rust code with the original C execution model. 
As a result, these approaches produce significantly inflated code size (e.g., RLOC increases exceeding 140\%) and extensive use of unsafe constructs (around 78\% on average), as shown in Table~\ref{tab: combined-project-results}. 
Moreover, when projects depend on external or system libraries, FFI-based translation requires complex hybrid builds between Rust and C, including manual configuration of modules, symbol resolution, and enabling unstable features, substantially increasing the burden on developers. 
In addition, the generated code largely preserves C-style memory management and control flow, yielding non-idiomatic Rust implementations that do not leverage native Rust abstractions. 
These results indicate that while rule-based methods can preserve behavioral equivalence, their semantic correctness reflects C-like execution rather than safe, idiomatic, and maintainable Rust code, limiting their practical applicability.

\saferrust attempts to address these issues by applying LLM-based refactoring to \ctorust outputs, but its effectiveness remains limited in project-level scenarios. 
Since the initial translation already introduces substantial unsafe code, expanded low-level logic, and tightly coupled dependencies, refactoring must operate on structurally complex, potentially inconsistent code. 
At the project level, this process requires repeatedly rewriting individual fragments, followed by compilation and linking, thereby significantly increasing repair costs and latency. 
More importantly, post-hoc refactoring cannot fundamentally resolve structural inconsistencies, such as mismatched interfaces, implicit dependencies, or redundant symbol definitions, that are introduced during the initial translation. 
As a result, \saferrust exhibits degraded $SemCor$ (30.07\%) and only moderate improvements in safety. 

In contrast, \tool achieves a fundamentally different trade-off between correctness, safety, and code quality. 
While its $SemCor$ is slightly lower than that of \ctorust, it substantially reduces unsafe usage (from 77.75\% to 8.48\% on average) and overall code size (achieving a negative code increment of $-6.76\%$). 
This improvement stems from its program-structure-aware design, which avoids direct reliance on C FFI and instead leverages Rust-native abstractions and explicit dependency modeling. 
By translating programs in a dependency-aware order and enforcing consistency of global symbols and interfaces, \tool eliminates redundant code and prevents structural inconsistencies that commonly arise in project-level translation. 
Furthermore, although \tool produces a moderate number of \texttt{Clippy} warnings, these are primarily stylistic (e.g., naming conventions) rather than functional issues, and the near-zero error rate confirms the reliability of the generated code. 
Overall, the results demonstrate that incorporating program structure and dependency information is essential for achieving a better balance between correctness, safety, and maintainability in project-level C-to-Rust translation.

\begin{figure}[ht]
  \centering
  \includegraphics[]{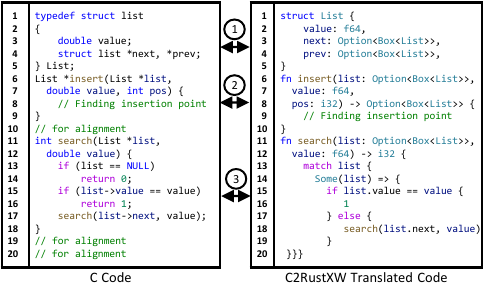}
  \caption{Doubly linked list program translated by \tool. (GitHub/doubly\_linked\_list.c).}
  \Description{This is the doubly linked list program example translated by \tool.}
  \label{fig: doublylinkedlist-xw}
\end{figure}

\subsection{RQ3: Case study}

To better understand the strengths and limitations of different approaches, we conduct case studies on representative C programs that expose common challenges in C-to-Rust translation, including array bounds handling and pointer-based data structures. We focus on two representative cases and analyze how different approaches behave under these scenarios.

\noindent\textbf{Case 1: Array handling and boundary safety.}
We first examine a counting sort implementation from CodeNet, where correct array indexing is critical. Rule-based approaches such as \ctorust and \genctorust translate the program into unsafe Rust code that preserves the original C-style array manipulation. Although these translations are syntactically correct, they lead to runtime errors due to out-of-bounds access in loop conditions, reflecting the lack of proper bounds reasoning. In contrast, \pltranslation generates more idiomatic Rust code, but may introduce inconsistencies in function interfaces and parameter usage due to the absence of structural constraints. \tool addresses these issues by leveraging program-structure-aware analysis: it enforces function interface consistency and replaces C-style array operations with Rust-native abstractions (e.g., vector length checks), effectively preventing out-of-bounds errors while maintaining concise code.

\noindent\textbf{Case 2: Pointer-based data structures.}
We further analyze a doubly linked list implementation from GitHub, which requires careful handling of pointers and memory ownership. Rule-based approaches again preserve raw pointer semantics, yielding unsafe Rust code that may compile but still causes runtime panics due to incorrect pointer manipulation. \genctorust slightly improves type safety through pointer retyping, but does not resolve deeper structural inconsistencies in list operations. LLM-based approaches such as \pltranslation and \sactor struggle with this case, often producing syntactically incorrect code or failing to preserve the intended data structure semantics. In contrast, \tool successfully translates programs by mapping C pointers to Rust ownership constructs such as \texttt{Option} and \texttt{Box}, thereby ensuring memory safety and eliminating runtime errors. By explicitly modeling data dependencies and enforcing consistent interfaces across functions, \tool produces a correct and idiomatic implementation. By explicitly modeling data dependencies and enforcing consistent interfaces across functions, \tool produces a correct and idiomatic implementation. 
This case further demonstrates that incorporating program structure enables \tool to effectively resolve common translation failures and complements the quantitative findings.

\section{Discussion}

\subsection{Insights and Implications}

Our results across RQ1--RQ3 reveal several important insights into the strengths and limitations of existing C-to-Rust translation approaches, and highlight the role of program structure in enabling effective translation. 
Rule-based methods, such as \ctorust and \genctorust, are highly effective in preserving the original C semantics by directly mapping C constructs to Rust equivalents and relying on unsafe operations or FFI when necessary. 
However, this design has significant practical drawbacks: the generated code often exhibits substantial code size inflation, a heavy reliance on unsafe constructs, and poor adherence to Rust idioms, as evidenced by both file-level and project-level results and case studies. 
Furthermore, when external libraries are involved, FFI-based translation requires complex hybrid builds, increasing the burden on developers and limiting usability. 
These findings suggest that semantic preservation alone is insufficient and that code quality, safety, and idiomatic design must be considered as first-class objectives.

Approaches that extend rule-based translation with LLM- or agent-based refactoring, such as \sactor and \saferrust, aim to improve code quality but remain fundamentally limited. Our results show that these methods often exhibit low syntactic and semantic correctness, especially at the file level, while incurring substantial computational and engineering costs due to iterative rewriting, compilation, and integration. These challenges are exacerbated by large code size, extensive unsafe usage, and complex dependencies from rule-based outputs, making post-hoc refactoring unreliable. This highlights that repairing code without explicit structural guidance is insufficient to resolve translation inconsistencies. 
In contrast, purely LLM-based approaches like \pltranslation generate more concise Rust code by leveraging learned knowledge, but their effectiveness diminishes for larger or multi-file codebases, where handling long-range dependencies and cross-file interactions consistently is crucial. Without structural constraints, LLMs struggle to maintain interface consistency and produce complete outputs, often resulting in incomplete or inconsistent code. Even with repair mechanisms, their reliance on unstructured code text limits the ability to reason about program semantics effectively, hindering reliable recovery from deeper structural errors.

These observations highlight that the key limitation of existing approaches is the absence of an explicit program structure as a guiding signal. 
In contrast, \tool explicitly models and leverages program structure throughout the entire translation pipeline. 
By extracting and encoding global symbols, dependency relationships, and control- and data-flow information to all stages, \tool enables dependency-aware translation that preserves interface consistency and reduces structural errors. 
Moreover, by ensuring syntactic correctness and parseability through multi-stage repair, subsequent function-level, item-level, and semantic repair processes can operate on well-structured code. 
The integration of both static and dynamic structural information further enables effective semantic error localization and correction. 
As a result, \tool consistently achieves high syntactic and semantic correctness at both file and project levels while significantly improving code safety and conciseness.
Overall, these findings suggest that incorporating explicit program structure into both generation and repair is critical for reliable cross-language translation. 
The structure-to-prompt paradigm adopted in \tool demonstrates that LLMs can be significantly enhanced when guided by structured program information, providing a promising direction for future research in code translation and program transformation.

\subsection{Threats to Validity}

Our evaluation is subject to several threats to validity. 
For external validity, although we evaluate \tool on diverse datasets from CodeNet and GitHub as well as multiple real-world C projects, these programs may not fully capture the complexity of large-scale industrial codebases with deeply nested dependencies and system-level interactions. 
For internal validity, while \tool achieves perfect syntactic correctness at the project level, semantic correctness varies across projects, indicating that preserving behavior for highly interdependent programs remains challenging. 
Although differential testing is employed to validate semantic equivalence, the coverage of available test cases may be insufficient to expose all corner cases. 
For construct validity, although unsafe usage is significantly reduced, \texttt{Clippy} warnings remain relatively frequent, primarily due to the preservation of C-style naming conventions to maintain traceability. 
Future work could address these limitations by incorporating larger and more complex project datasets, improving semantic validation through enhanced static and dynamic analysis techniques, and introducing automated refactoring to generate more idiomatic Rust code.

\section{Related Work}

Program translation has been studied through a wide range of approaches, including rule-based, statistical, neural, and LLM-driven methods. Early work focused on rule-based and abstraction-driven translation~\cite{qiu1999programming, waters2002program}, emphasizing language-independent representations and systematic reimplementation strategies. Subsequent research explored statistical machine translation and neural models~\cite{karaivanov2014phrase, chen2018tree, drissi2018program}, which improved translation quality by learning mappings between programming languages. More recent approaches leverage intermediate representations and large-scale pretraining~\cite{transcoder-IR, huang2023program, liu2023syntax}, achieving strong performance on languages with similar abstractions such as Java, Python, and C++. However, these methods are less effective due to fundamental differences in memory safety models and the limited availability of high-quality parallel corpora.

Recent advances in LLM-based code generation have further improved the flexibility and capability of automated translation. Prior work explores techniques such as evaluation-driven generation~\cite{wang2023review}, compiler testing~\cite{gu2023llm}, hallucination analysis~\cite{liu2024exploring}, interactive workflows~\cite{fakhoury2024llm}, bias mitigation~\cite{huang2024bias}, retrieval augmentation~\cite{koziolek2024llm}, multi-agent collaboration~\cite{nunez2024autosafecoder}, and test-driven generation~\cite{mathews2024test, peng2025cweval}. While these approaches improve correctness and robustness through testing and feedback, they often lack explicit integration of program structure and static analysis, which is critical for ensuring structural consistency and dependency correctness in code translation tasks. In contrast, \tool incorporates program-structure-aware analysis, dependency-guided translation, and multi-stage syntactic and semantic repair into a unified framework. By explicitly modeling global symbols and program dependencies, \tool enables more reliable and scalable C-to-Rust translation across function-, file-, and project-level scenarios, producing code that is both syntactically correct and semantically consistent.

\section{Conclusion}

In this paper, we presented \tool, a program-structure-aware framework for C-to-Rust translation that integrates static program analysis with LLM-based generation and repair. 
\tool explicitly models multi-level program structure and transforms it into textual representations that are injected into LLM prompts, enabling dependency-aware translation and structure-guided repair within a unified pipeline. 
Experimental results demonstrate the effectiveness of \tool across both benchmark datasets and real-world projects: it achieves 100\% syntactic correctness on CodeNet and 97.78\% on GitHub at the file level, and perfect syntactic correctness with an average semantic correctness of 78.87\% at the project level, while significantly reducing code size and unsafe usage. 
These results show that \tool can produce concise, safe, and semantically consistent Rust code, providing a practical and scalable solution for reliable C-to-Rust migration.




\bibliographystyle{ACM-Reference-Format}
\bibliography{ASE-main-ref}

\end{document}